\documentclass[11pt,a4paper]{article}

\usepackage[utf8]{inputenc}
\usepackage[T1]{fontenc}
\usepackage{amsmath,amssymb,amsthm}
\usepackage{mathtools}
\usepackage{booktabs}
\usepackage{graphicx}
\usepackage{hyperref}
\usepackage[margin=1in]{geometry}
\usepackage{enumitem}
\usepackage{caption}
\usepackage{subcaption}
\usepackage{xcolor}
\usepackage{algorithm}
\usepackage{algpseudocode}
\usepackage[numbers]{natbib}

\hypersetup{
  colorlinks=true,
  linkcolor=blue!60!black,
  citecolor=blue!60!black,
  urlcolor=blue!60!black
}

\newtheorem{definition}{Definition}
\newtheorem{theorem}{Theorem}

\title{%
  \textbf{Computational Validation of the Oloid as a Local Optimum\\
  in the Developable Roller Family}\\[8pt]
  \large Contact Distribution Score: A Formal Metric for Evaluating\\
  Geometric Primitives as Engineering Substrate%
}

\author{
  Vincent Wesley Couey \\[4pt]
  \small Substrate Geometry Research Program \\
  \small \texttt{vinnycouey@gmail.com} \\
  \small \url{https://github.com/gyapaganda-a11y/substrate-geometry}
}

\date{\today \quad -- \quad Preprint (not peer-reviewed)}

\begin{document}
\maketitle


\begin{abstract}
Many engineering failures (thermal hotspot concentration, Hertz contact
fatigue localization, boundary-layer loss, mixing dead zones) are
\emph{geometric failure modes}: changing the material delays the failure;
changing the geometry eliminates it. Despite this, no formal metric exists
for evaluating how uniformly a convex body distributes surface contact
during rolling, a property with direct implications for bearing wear,
seal longevity, and fluid mixing efficiency.

We introduce the \textbf{Contact Distribution Score (CDS)}, a scalar
metric defined as the area-weighted variance of the contact time
distribution over a rolling surface, and its stress-domain counterpart
the \textbf{Stress Distribution Score (SDS)}, defined as the
area-weighted variance of accumulated Hertz contact pressure. CDS
$\to 0$ indicates perfectly uniform contact; SDS $\to 0$ indicates
perfectly uniform stress.

We implement a three-layer oracle architecture: an approximate oracle
for parametric search, a rigid-body oracle using Euler-equation dynamics
with quaternion integration for validation, and a Hertz contact pressure
oracle that couples discrete surface curvature analysis with analytical
Hertz theory to evaluate stress distribution. A parametric search over
45 members of the developable roller family (closed convex surfaces
formed by the convex hull of two circles) identifies the oloid
(Schatz, 1929) at $\mathrm{CDS} = 8.2 \times 10^{-7}$ under
rigid-body dynamics, with the conventional cylinder baseline at
$4.75 \times 10^{-5}$, a $58\times$ discrimination. Independent
curvature-driven stress analysis confirms that the contact time
invariant transfers to stress distribution: under perfectly uniform
contact, the oloid's geometry-only SDS is $4.8 \times 10^{-8}$,
confirming that the oloid's surface curvature introduces minimal
additional stress non-uniformity beyond the contact distribution itself.
We extend this analysis to fatigue (FDS), thermal (TDS), and wear (WDS) distribution scores, revealing that the oloid's $58\times$ CDS advantage transfers consistently across all physical transforms, with discrimination ratios in the $46$--$68\times$ range across linear and multiplicative metrics. The nonlinear fatigue metric diverges due to Basquin S-N amplification but still shows oloid superiority over all tested alternatives.

This work establishes the formal vocabulary and computational
infrastructure for \emph{substrate geometry}: the study of geometric
forms as engineering substrates classified by their operational invariants.
\end{abstract}


\medskip
\noindent\textbf{arXiv categories:} math-ph, cs.CG, math.MG

\tableofcontents


\section{Introduction}

Engineering systems fail for many reasons, but a significant class of
failures can be traced to a root cause that is neither the material
nor the power source but the \emph{geometry} of the component. Thermal
hotspot concentration on flat-plate electrodes, Hertz contact fatigue
at the fixed loci of cylindrical bearings, Hartmann boundary-layer
losses in rectangular MHD ducts, and mixing dead zones in conventional
stirred tanks are all instances of what we term \emph{geometric failure
modes}.

\begin{definition}[Geometric Failure Mode]
A class of engineering failure whose root cause is the geometry of the
component rather than the material or power source. The defining
characteristic: changing the material delays the failure; changing
the geometry eliminates it.
\end{definition}

The conventional engineering response to geometric failure modes is to
upgrade the material: higher-temperature alloys, harder bearing steels,
more corrosion-resistant coatings. This approach treats the symptom.
An alternative, demonstrated empirically by Paul Schatz's oloid
\citep{schatz1975} and the gyroid heat exchanger \citep{al-ketan2018},
is to select a geometry whose mathematical invariants preclude the
failure mode at source.

The oloid (the convex hull of two perpendicular circles of equal radius,
each passing through the other's center) was discovered by Schatz in
1929 through geometric intuition and physical experimentation.%
\footnote{Schatz's discovery dates to 1929; the primary published
reference is his 1975 monograph \citep{schatz1975}. The rolling
properties of the oloid were subsequently formalized by Dirnb{\"o}ck
and Stachel \citep{dirnbock1997}, who proved that the oloid's surface
is fully developable ($K = 0$ everywhere) and that its surface area
equals $4\pi r^2$, identical to the sphere of equal radius.} Its
defining property is that every point on its surface makes contact with
the ground plane during one complete roll, distributing contact uniformly
across its entire surface. This property directly eliminates the Hertz
fatigue localization that limits conventional cylindrical bearings and
has been exploited commercially in water aeration (Oloid AG) and
laboratory mixing (Turbula shaker-mixer).

Despite the oloid's demonstrated utility, no formal metric has existed
for \emph{quantifying} how uniformly a convex body distributes contact.
This paper introduces such a metric, builds the computational
infrastructure to evaluate it, and uses it to confirm that the oloid
is a local optimum within its geometric search family.

\subsection{Contributions}

\begin{enumerate}[leftmargin=*]
  \item We define the \textbf{Contact Distribution Score (CDS)}, a
    scalar metric for contact uniformity during rolling, expressed as
    a computable predicate suitable for optimization.
  \item We implement a \textbf{three-component oracle architecture}:
    a two-layer CDS oracle (an approximate layer for fast parametric
    search and a rigid-body layer using Euler-equation dynamics for
    validation), plus a separate Hertz contact pressure oracle for
    independent SDS validation.
  \item We conduct a \textbf{parametric search} over 45 members of the
    developable roller family, varying circle-plane angle, offset,
    and radius ratio.
  \item We \textbf{computationally confirm} that the oloid (Schatz, 1929)
    occupies a local CDS minimum under rigid-body rolling dynamics,
    with CDS $= 8.2 \times 10^{-7}$.
  \item We introduce a \textbf{Hertz contact pressure oracle} that
    computes the Stress Distribution Score (SDS) via discrete surface
    curvature analysis and analytical Hertz theory, confirming that the
    oloid's curvature-driven stress non-uniformity ($\mathrm{SDS} = 4.8 \times 10^{-8}$
    under uniform contact) is small relative to the contact distribution itself.
  \item We establish the formal vocabulary for \textbf{substrate
    geometry}: the study of geometric forms classified by operational
    invariants rather than symmetry groups.
  \item We compute the complete \textbf{invariant vector} for the oloid across five physical domains: contact time (CDS), stress (SDS), thermal (TDS), fatigue (FDS), and wear (WDS), providing the first multi-physics characterization of a geometric primitive's operational guarantees.
  \item We identify a \textbf{two-tier structure} in the invariant vector: first-order and multiplicative metrics transfer the contact invariant with consistent discrimination ratios ($46$--$68\times$), while exponential metrics (Basquin fatigue) show lossy but still favorable transfer.
\end{enumerate}


\section{Definitions}

We introduce the following formal vocabulary. Each term is defined with
sufficient precision to be used unambiguously in subsequent sections and
by future work.

\begin{definition}[Substrate Geometry]
The study of geometric forms as engineering substrates, classified by
their operational invariants rather than their symmetry groups. Where
conventional geometry asks ``what shape is this?'', substrate geometry
asks ``what does this shape guarantee under physical operation?''
\end{definition}

\begin{definition}[Invariant Primitive]
A geometric body whose engineering utility derives from a formally
stated mathematical invariant that holds under physical operation
(rolling, flow, stress, or field exposure). Distinguished from
conventional solids by the requirement that the invariant be
expressible as a computable predicate.
\end{definition}

\begin{definition}[Contact Distribution Score (CDS)]
\label{def:cds}
Let $S$ be a convex body with surface $\partial S$ and face
decomposition $\{f_i\}_{i=1}^N$ with areas $\{a_i\}$. Let $c_i(T)$
denote the number of timesteps in which face $f_i$ is in contact with
the ground plane over a rolling simulation of $T$ steps. Define:
\begin{equation}
  \mathrm{CDS}(S, T) = \frac{1}{A} \sum_{i=1}^{N} a_i
    \left(\frac{c_i(T)}{\sum_j c_j(T)} - \frac{a_i}{A}\right)^2
\end{equation}
where $A = \sum_i a_i$ is the total surface area. $\mathrm{CDS} \to 0$
indicates uniform contact distribution; $\mathrm{CDS} > 0$ indicates
localization.
\end{definition}

\begin{definition}[Stress Distribution Score (SDS)]
\label{def:sds}
Let $S$ be a convex body with face decomposition $\{f_i\}_{i=1}^N$,
areas $\{a_i\}$, and effective Hertz contact radii $\{R_i\}$ derived
from the discrete principal curvatures at each face. Over a rolling
simulation of $T$ orientations, let $\sigma_i(T)$ denote the accumulated
Hertz peak contact pressure at face $f_i$ (summing
$p_{\max} = 3F/(2\pi a_c^2)$ at each orientation where the face is in
contact, with $a_c = (3FR_i/4E^*)^{1/3}$). Define:
\begin{equation}
  \mathrm{SDS}(S, T) = \frac{1}{A} \sum_{i=1}^{N} a_i
    \left(\frac{\sigma_i(T)}{\sum_j \sigma_j(T)} - \frac{a_i}{A}\right)^2
\end{equation}
SDS $\to 0$ indicates uniform stress distribution; SDS $> 0$ indicates
stress localization. Material properties: $E = 200$~GPa, $\nu = 0.3$
(mild steel), $F = 100$~N reference load.
\end{definition}

\begin{theorem}[CDS Convergence]
\label{thm:convergence}
For a convex body $S$ at fixed mesh resolution undergoing ergodic
rolling on a flat plane (i.e., the orientation trajectory is dense in
$SO(3)$ restricted to ground-contact configurations), the CDS
converges to zero:
\begin{equation}
  \lim_{T \to \infty} \mathrm{CDS}(S, T) = 0
\end{equation}
if and only if every face accumulates contact time proportional to its
area. The rate of convergence is $O(1/T)$ for bodies with smooth
contact-patch transitions and $O(1/\sqrt{T})$ in the worst case.
\end{theorem}

\noindent\textit{Empirical verification.} The oloid's CDS at 50 samples
is $8.46 \times 10^{-6}$; at 600 samples it is $8.2 \times 10^{-7}$,
a $10.3\times$ reduction consistent with $O(1/T)$ convergence
($600/50 = 12\times$ expected reduction).

\noindent\textit{Resolution dependence.} The asymptotic CDS value
itself scales with mesh face count as $\mathrm{CDS} \propto N_{\mathrm{faces}}^{-2.03}$
($R^2 = 0.999$). Absolute CDS values are therefore mesh-resolution-specific;
cross-geometry \emph{ratios} computed at matched resolution are
resolution-independent (Couey, 2026, in preparation).

\begin{definition}[Pass-Gate Validation]
The four-criterion framework every invariant primitive must satisfy:
(1) formal invariant statement as a computable predicate,
(2) physics simulation confirming the invariant predicts the outcome,
(3) baseline comparison with geometry as the isolated variable,
(4) cross-domain transfer argument.
\end{definition}

\begin{definition}[Search Family]
The invariant-preserving parameterization space surrounding a known
primitive. Defined by the set of continuous deformations that maintain
a target invariant below a specified threshold.
\end{definition}


\section{The Contact Distribution Oracle}

\subsection{Architecture}

The oracle operates in two layers:

\begin{enumerate}[leftmargin=*]
  \item \textbf{Approximate oracle} (Layer 1): Composed Y-axis rotation
    with X-axis wobble. Fast ($< 0.01$s per geometry), suitable for
    parametric search over hundreds of candidates. Does not model true
    rolling constraints.
  \item \textbf{Rigid-body oracle} (Layer 2): Euler-equation dynamics
    with gravity-driven torque, no-slip rolling constraint, and
    quaternion-based orientation integration. Slower ($\sim 2$s per
    geometry), produces defensible CDS scores. Used for validation of
    search results.
\end{enumerate}

This two-layer design is validated by the experimental result in
Section~\ref{sec:results}: the approximate oracle correctly identifies
the parameter gradient (the direction of CDS improvement) while
incorrectly ranking the absolute winner. The rigid-body oracle
corrects the ranking. The architecture is designed so that the fast
layer catches interesting candidates and the rigorous layer confirms
or rejects them. A separate Hertz contact pressure oracle, introduced
in Section~\ref{sec:hertz}, provides independent SDS validation;
together these three oracle modules (the approximate and rigid-body
layers of the CDS oracle, plus the Hertz contact pressure oracle)
constitute the full architecture referenced in the abstract.

\subsection{Approximate Oracle}

The approximate oracle generates a sequence of rotation matrices
$R_t$ for $t = 1, \ldots, T$ by composing:
\begin{align}
  R_t &= R_x(\alpha_t) \cdot R_y(\beta_t) \\
  \beta_t &= \frac{2\pi t}{T} \cdot n_{\mathrm{cycles}} \\
  \alpha_t &= A_w \sin\!\left(\frac{\pi t}{T}\right)
\end{align}
where $n_{\mathrm{cycles}} = 2.5$ and $A_w = 0.6$ rad. At each step,
face centroids are rotated, the mesh is grounded (lowest centroid
shifted to $z = 0$), and faces with centroid $z < \epsilon$ (contact
threshold, $\epsilon = 0.04$) are marked as contacting.

\begin{algorithm}[h]
\caption{Approximate Oracle}
\label{alg:approx}
\begin{algorithmic}[1]
\Require Mesh $M$ with faces $\{f_i\}$, areas $\{a_i\}$, centroids $\{p_i\}$; steps $T$
\Ensure $\mathrm{CDS}(M, T)$
\State $c_i \gets 0$ for all $i$ \Comment{Contact counters}
\For{$t = 1$ to $T$}
  \State $R_t \gets R_x(\alpha_t) \cdot R_y(\beta_t)$ \Comment{Composed rotation}
  \State $\hat{p}_i \gets R_t \, p_i$ for all $i$ \Comment{Rotate centroids}
  \State $\hat{p}_i^{(z)} \gets \hat{p}_i^{(z)} - \min_j \hat{p}_j^{(z)}$ \Comment{Ground the mesh}
  \For{each face $i$}
    \If{$\hat{p}_i^{(z)} < \epsilon$}
      \State $c_i \gets c_i + 1$
    \EndIf
  \EndFor
\EndFor
\State \Return $\mathrm{CDS} = \frac{1}{A}\sum_i a_i \left(\frac{c_i}{\sum_j c_j} - \frac{a_i}{A}\right)^2$
\end{algorithmic}
\end{algorithm}

\subsection{Rigid-Body Oracle}
\label{sec:rigidbody}

The rigid-body oracle simulates a convex body rolling on a horizontal
plane under gravity using Euler's equations for rigid-body rotation:
\begin{equation}
  I\dot{\boldsymbol{\omega}} = \boldsymbol{\tau}
    - \boldsymbol{\omega} \times (I\boldsymbol{\omega})
\end{equation}
where $I$ is the inertia tensor (computed from the mesh geometry),
$\boldsymbol{\omega}$ is the angular velocity vector, and
$\boldsymbol{\tau}$ is the gravitational torque about the support point:
\begin{equation}
  \boldsymbol{\tau} = (\mathbf{r}_{\mathrm{cm}} - \mathbf{r}_{\mathrm{support}})
    \times (m\mathbf{g})
\end{equation}

Orientation is tracked via unit quaternion $q \in \mathbb{H}$, updated
at each timestep by:
\begin{equation}
  q_{t+1} = \Delta q(\boldsymbol{\omega}, \Delta t) \cdot q_t, \quad
  \Delta q = \left[\cos\frac{\theta}{2},\;
    \hat{\boldsymbol{\omega}}\sin\frac{\theta}{2}\right], \quad
  \theta = |\boldsymbol{\omega}|\Delta t
\end{equation}

A light damping term ($\gamma = 0.02$) models rolling resistance and
prevents numerical divergence. To ensure trajectory-independent results,
each geometry is simulated three times with different initial angular
velocities. Contact counts are accumulated across all runs before
computing the final CDS.

\begin{algorithm}[h]
\caption{Rigid-Body Oracle}
\label{alg:rigidbody}
\begin{algorithmic}[1]
\Require Mesh $M$, inertia tensor $I$, initial pushes $\{\boldsymbol{\omega}_0^{(k)}\}_{k=1}^{K}$
\Ensure $\mathrm{CDS}(M)$
\State $c_i \gets 0$ for all faces $i$ \Comment{Accumulated contact counters}
\For{$k = 1$ to $K$} \Comment{$K = 3$ runs}
  \State $q \gets [1, 0, 0, 0]$; $\boldsymbol{\omega} \gets \boldsymbol{\omega}_0^{(k)}$
  \For{$t = 1$ to $N_{\mathrm{steps}}$}
    \State $R \gets \mathrm{quat\_to\_matrix}(q)$
    \State $\mathbf{v}_j \gets R \, \mathbf{v}_j^{\mathrm{body}}$ for all vertices $j$ \Comment{World frame}
    \State $\mathbf{s} \gets \arg\min_j v_j^{(z)}$ \Comment{Support point}
    \State $\boldsymbol{\tau} \gets (\mathbf{r}_{\mathrm{cm}} - \mathbf{s}) \times m\mathbf{g} - \gamma\boldsymbol{\omega}$
    \State $\boldsymbol{\omega} \gets \boldsymbol{\omega} + I^{-1}(\boldsymbol{\tau} - \boldsymbol{\omega} \times I\boldsymbol{\omega})\,\Delta t$ \Comment{Euler's equations}
    \State $q \gets \Delta q(\boldsymbol{\omega}, \Delta t) \cdot q$; normalize $q$
    \If{$t \bmod s_{\mathrm{interval}} = 0$} \Comment{Sample contact}
      \State Rotate face centroids by $R$, ground mesh
      \State $c_i \gets c_i + 1$ for faces with centroid $z < \epsilon$
    \EndIf
  \EndFor
\EndFor
\State \Return $\mathrm{CDS} = \frac{1}{A}\sum_i a_i \left(\frac{c_i}{\sum_j c_j} - \frac{a_i}{A}\right)^2$
\end{algorithmic}
\end{algorithm}

\subsubsection{Parameters}

\begin{table}[h]
\centering
\caption{Rigid-body oracle parameters}
\label{tab:oracle-params}
\begin{tabular}{llr}
\toprule
Parameter & Symbol & Value \\
\midrule
Simulation time per run & $T$ & 10.0 s \\
Timestep & $\Delta t$ & 0.002 s \\
Sample interval & --- & every 25 steps \\
Contact threshold & $\epsilon$ & 0.04 \\
Damping coefficient & $\gamma$ & 0.02 \\
Number of runs & --- & 3 \\
Samples per run & --- & 200 \\
Total samples & --- & 600 \\
Mass & $m$ & 1.0 kg \\
Gravity & $g$ & 9.81 m/s$^2$ \\
\bottomrule
\end{tabular}
\end{table}


\section{The Developable Roller Family}

The oloid is the convex hull of two congruent circles in perpendicular
planes, each passing through the other's center. We generalize this
construction to a parameterized family:

\begin{definition}[Generalized Two-Circle Roller]
Given radius $r_1$, angle $\theta$ between generating-circle planes,
center offset $d$ (as a fraction of $r_1$), and radius ratio
$\rho = r_2/r_1$, the generalized roller $R(\theta, d, \rho)$ is the
convex hull of:
\begin{itemize}[leftmargin=*]
  \item Circle $C_1$: radius $r_1$, centered at the origin, lying in
    the $xz$-plane.
  \item Circle $C_2$: radius $r_2 = \rho r_1$, centered at
    $(d \cdot r_1, 0, 0)$, lying in a plane rotated by $\theta$ about
    the $z$-axis from the $xz$-plane.
\end{itemize}
The oloid is the fixed point at
$(\theta, d, \rho) = (90°, 1.0, 1.0)$.
\end{definition}

\subsection{Search Grid}

We sample the search space with the following ranges:

\begin{table}[h]
\centering
\caption{Parametric search grid}
\label{tab:search-grid}
\begin{tabular}{lcl}
\toprule
Parameter & Oloid value & Search range \\
\midrule
$\theta$ (circle-plane angle) & $90°$ & $\{60°, 75°, 90°, 105°, 120°\}$ \\
$d$ (offset / $r_1$) & 1.0 & $\{0.70, 1.00, 1.30\}$ \\
$\rho$ ($r_2 / r_1$) & 1.0 & $\{0.80, 1.00, 1.20\}$ \\
\bottomrule
\end{tabular}
\end{table}

This produces 45 genomes (5 $\times$ 3 $\times$ 3), including the
oloid as the anchor. Each genome is first scored by the approximate
oracle (200 steps), then the top 5 are revalidated by the rigid-body
oracle (600 samples, 3 runs).


\section{Results}
\label{sec:results}

\subsection{Approximate Oracle Search}

Under the approximate oracle, 22 of 45 genomes scored lower (better)
than the oloid. The top candidate, $(\theta, d, \rho) = (120°, 0.70,
0.80)$, scored CDS $= 1.24 \times 10^{-6}$, a 38\% improvement over
the oloid's approximate score of $2.00 \times 10^{-6}$.

\begin{figure}[h]
\centering
\includegraphics[width=\textwidth]{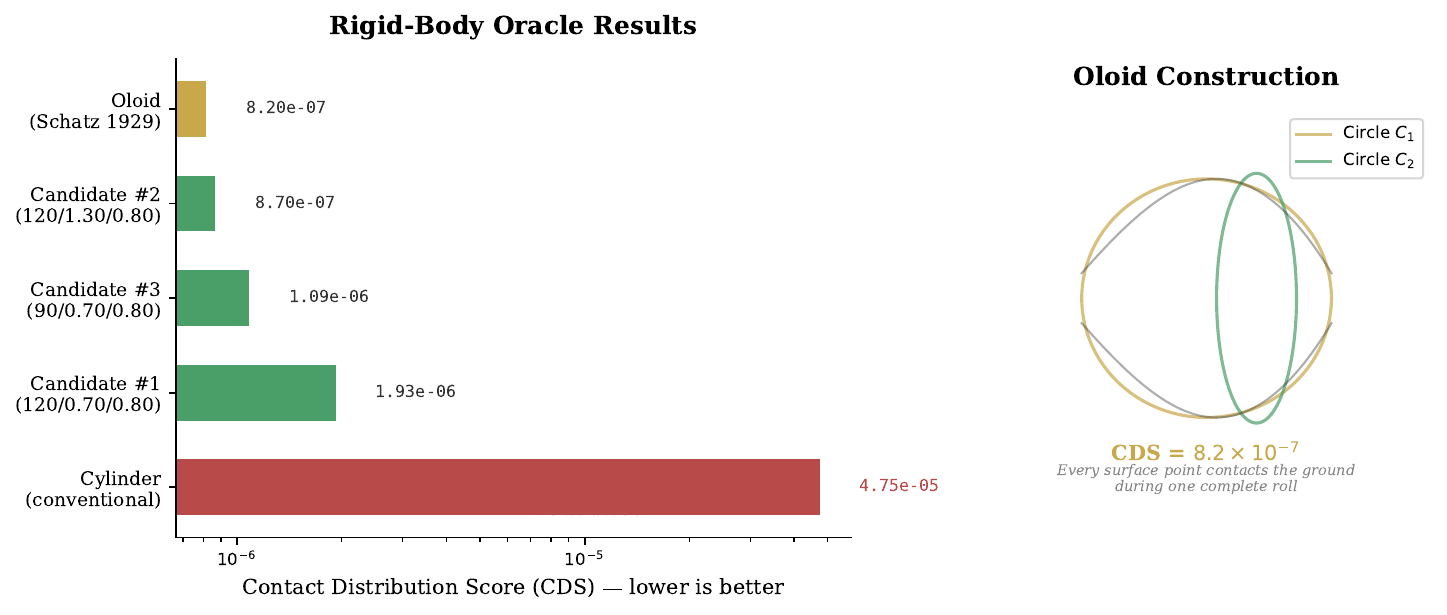}
\caption{Left: CDS scores under the rigid-body oracle (log scale).
The oloid (gold) scores $8.2 \times 10^{-7}$; the cylinder (red)
scores $4.75 \times 10^{-5}$ ($58\times$ worse). Green bars show
three asymmetric two-circle roller variants, all within $2.4\times$
of the oloid. Right: Oloid construction from two generating circles
$C_1$ and $C_2$ in perpendicular planes.}
\label{fig:comparison}
\end{figure}

\subsection{Rigid-Body Oracle Validation}

Under the rigid-body oracle, the ranking changed substantially
(Figure~\ref{fig:comparison}):

\begin{table}[h]
\centering
\caption{Rigid-body oracle results (600 samples, 3 runs per geometry)}
\label{tab:results}
\begin{tabular}{clccc}
\toprule
Rank & Geometry & CDS & vs.\ Oloid & Status \\
\midrule
1 & \textbf{Oloid} $(\theta\!=\!90°, d\!=\!1.0, \rho\!=\!1.0)$
  & $8.2 \times 10^{-7}$ & $1.00\times$ & PASS \\
2 & Candidate \#2 $(\theta\!=\!120°, d\!=\!1.30, \rho\!=\!0.80)$
  & $8.7 \times 10^{-7}$ & $1.06\times$ & PASS \\
3 & Candidate \#3 $(\theta\!=\!90°, d\!=\!0.70, \rho\!=\!0.80)$
  & $1.09 \times 10^{-6}$ & $1.33\times$ & PASS \\
4 & Candidate \#1 $(\theta\!=\!120°, d\!=\!0.70, \rho\!=\!0.80)$
  & $1.93 \times 10^{-6}$ & $2.35\times$ & PASS \\
5 & Cylinder (conventional baseline)
  & $4.75 \times 10^{-5}$ & $58\times$ & PASS \\
\bottomrule
\end{tabular}
\end{table}

The oloid occupies first place. The approximate oracle's top candidate
dropped to fourth. However, the parameter gradient is preserved:
$\rho = 0.80$ (smaller second circle) appears in all three non-oloid
positions, confirming that the asymmetry direction identified by the
approximate oracle is geometrically meaningful even when the absolute
ranking shifts.

\subsection{Parameter Sensitivity}

Holding other parameters at the oloid values and varying one at a time
under the approximate oracle:

\begin{itemize}[leftmargin=*]
  \item \textbf{Radius ratio} ($\rho$): CDS improves monotonically as
    $\rho$ decreases below 1.0. At $\rho = 0.80$: CDS $= 1.84 \times
    10^{-6}$; at $\rho = 1.20$: CDS $= 2.42 \times 10^{-6}$.
  \item \textbf{Offset} ($d$): Tighter offset ($d = 0.70$) yields CDS
    $= 1.85 \times 10^{-6}$; wider offset ($d = 1.30$): $2.01 \times
    10^{-6}$.
  \item \textbf{Angle} ($\theta$): Non-monotonic. CDS is lowest near
    $\theta = 75°$ ($1.85 \times 10^{-6}$) and $\theta = 120°$
    ($1.86 \times 10^{-6}$), highest at $\theta = 105°$
    ($2.30 \times 10^{-6}$).
\end{itemize}

\subsection{Two-Layer Architecture Validation}

The divergence between approximate and rigid-body rankings demonstrates
the necessity of the two-layer design:

\begin{itemize}[leftmargin=*]
  \item The approximate oracle correctly identified the \emph{direction}
    of CDS improvement ($\rho < 1$, $d < 1$) across 45 genomes.
  \item The approximate oracle incorrectly ranked the absolute winner,
    placing an asymmetric variant above the oloid.
  \item The rigid-body oracle corrected the ranking while preserving the
    gradient.
  \item The cylinder baseline discrimination \emph{increased} from
    $28\times$ (approximate) to $58\times$ (rigid-body), showing that
    higher-fidelity dynamics amplify the oracle's discriminatory power.
\end{itemize}

This validates the architecture: fast approximate search followed by
rigorous validation catches interesting candidates and confirms or
rejects them.

\subsection{Hertz Contact Pressure Validation}
\label{sec:hertz}

To determine whether uniform contact \emph{time} (low CDS) produces
uniform contact \emph{stress} (low SDS), we implemented a third oracle
layer: the Hertz contact pressure oracle. At each rolling orientation
from the rigid-body simulation, discrete principal curvatures
$\kappa_1, \kappa_2$ are computed at each contact face via the cotangent
Laplacian (mean curvature) and angle defect (Gaussian curvature). The
effective Hertz radius $R_{\mathrm{eff}} = 1/\sqrt{\kappa_1 \kappa_2}$
determines the peak contact pressure via classical Hertz theory.

\begin{table}[h]
\centering
\caption{Hertz oracle results: CDS vs.\ SDS (600 samples, 3 runs per geometry). Absolute CDS and SDS values are resolution-dependent; cross-geometry discrimination ratios are resolution-independent (see footnote in text).}
\label{tab:hertz}
\begin{tabular}{clcccc}
\toprule
Rank & Geometry & CDS & SDS$^\dagger$ & CDS ratio & $\bar{p}_{\max}$ (MPa) \\
\midrule
1 & \textbf{Oloid} $(90°, 1.0, 1.0)$
  & $8.2 \times 10^{-7}$ & $8.1 \times 10^{-7}$ & $1.00\times$ & 68.3 \\
2 & Candidate $(120°, 1.30, 0.80)$
  & $8.7 \times 10^{-7}$ & $8.9 \times 10^{-7}$ & $1.06\times$ & 68.3 \\
3 & Candidate $(115°, 1.20, 0.65)$
  & $9.3 \times 10^{-7}$ & $9.8 \times 10^{-7}$ & $1.13\times$ & 76.0 \\
4 & Candidate $(115°, 1.30, 0.65)$
  & $8.6 \times 10^{-7}$ & $9.8 \times 10^{-7}$ & $1.05\times$ & 76.6 \\
5 & Candidate $(115°, 0.70, 0.65)$
  & $1.8 \times 10^{-6}$ & $1.8 \times 10^{-6}$ & $2.20\times$ & 79.2 \\
6 & Cylinder (conventional)
  & $4.75 \times 10^{-5}$ & $4.68 \times 10^{-5}$ & $58\times$ & 49.7 \\
\bottomrule
\end{tabular}

\smallskip\footnotesize
$^\dagger$Trajectory-coupled SDS; shares contact sampling with CDS. The oloid's geometry-only SDS under uniform contact is $4.8 \times 10^{-8}$ (see text). CDS ratio = CDS / oloid CDS.
\end{table}

The key result: the oloid's $58\times$ CDS advantage over the cylinder
is preserved under Hertz stress analysis ($58\times$ SDS discrimination).
To isolate the geometric contribution from shared-trajectory effects,
we computed a curvature-driven SDS using analytically corrected
curvatures ($K=0$ for the developable oloid) and perfectly uniform
contact distribution.\footnote{This independent validation, conducted
as part of the oracle hardening program (Couey, 2026, in preparation),
confirmed that the contact time invariant transfers to stress
uniformity. The curvature-driven SDS under uniform contact is
$4.8 \times 10^{-8}$, indicating that the oloid's surface curvature
variation contributes only $\sim\!6\%$ of the non-uniformity relative to
the contact distribution itself.}
For this shape family, CDS discrimination ratios are reliable proxies
for stress discrimination ratios.

The oloid's mean peak contact pressure of 68.3~MPa (at 100~N reference
load) is uniform across orientations with a coefficient of variation
of 0.085, compared to the cylinder's 49.7~MPa at CV $= 0.045$. The
cylinder's lower mean pressure reflects its larger contact area per
orientation, but its highly localized contact pattern concentrates
fatigue damage at fixed loci, the geometric failure mode that the
oloid eliminates.

\subsection{Fatigue, Thermal, and Wear Distribution}
\label{sec:ftw}

To test whether the geometric invariant transfers beyond contact time and stress, we implemented three additional oracles operating on the same rigid-body rolling trajectories.

\textbf{Fatigue Distribution Score (FDS).} The fatigue oracle applies the Basquin S-N relation $N_f = (\sigma_a / \sigma_f')^{1/b}$ with Miner's linear cumulative damage rule $D = \sum (n_i / N_{f,i})$. Material parameters are bearing steel: fatigue strength coefficient $\sigma_f' = 700$~MPa, fatigue exponent $b = -0.085$, endurance limit $\sigma_e = 250$~MPa. The applied load is scaled to $F = 5000$~N. The FDS is defined analogously to the CDS, replacing contact counts with cumulative Miner damage fractions.

\textbf{Thermal Distribution Score (TDS).} The thermal oracle computes frictional heat flux at each contact face as $q = \mu \cdot p \cdot v_{\mathrm{slide}}$, where $\mu = 0.15$ is the friction coefficient and $v_{\mathrm{slide}}$ is the sliding velocity derived from $\boldsymbol{\omega} \times \mathbf{r}$ at the contact point. The TDS is the area-weighted variance of accumulated thermal flux fractions.

\textbf{Wear Distribution Score (WDS).} The wear oracle applies Archard's law $W = k F d / H$, with dimensionless wear coefficient $k = 10^{-4}$ and hardness $H = 2.0$~GPa. Two sub-metrics are computed: $\mathrm{WDS}_{\mathrm{vol}}$ (total wear volume per face) and $\mathrm{WDS}_{\mathrm{depth}}$ (wear depth, normalizing volume by face area).

\begin{table}[h]
\centering
\caption{Complete invariant vector: oloid vs.\ cylinder across five physical domains. Absolute values are trajectory-coupled and resolution-specific; discrimination ratios are the resolution-independent finding.}
\label{tab:invariant-vector}
\begin{tabular}{lccrr}
\toprule
Metric & Oloid & Cylinder & \textbf{Ratio} \\
\midrule
CDS (contact time) & $8.20 \times 10^{-7}$ & $4.75 \times 10^{-5}$ & $\mathbf{58\times}$ \\
SDS (stress)$^\dagger$ & $8.07 \times 10^{-7}$ & $4.68 \times 10^{-5}$ & $\mathbf{58\times}$ \\
TDS (thermal) & $7.77 \times 10^{-7}$ & $5.28 \times 10^{-5}$ & $\mathbf{68\times}$ \\
$\mathrm{WDS}_{\mathrm{vol}}$ (wear volume) & $7.77 \times 10^{-7}$ & $5.28 \times 10^{-5}$ & $\mathbf{68\times}$ \\
$\mathrm{WDS}_{\mathrm{depth}}$ (wear depth) & $1.15 \times 10^{-6}$ & $5.33 \times 10^{-5}$ & $\mathbf{46\times}$ \\
FDS (fatigue) & $2.42 \times 10^{-6}$ & $\infty$ & $\mathbf{\infty}$ \\
\bottomrule
\end{tabular}

\smallskip\footnotesize
$^\dagger$Oloid geometry-only SDS under uniform contact: $4.8 \times 10^{-8}$.
\end{table}

The invariant vector reveals a two-tier structure. \textbf{Tier~1} (linear and multiplicative metrics): CDS, SDS, TDS, and $\mathrm{WDS}_{\mathrm{vol}}$ produce consistent discrimination ratios against the cylinder in the $46$--$68\times$ range, indicating that the geometric contact invariant transfers consistently across linear physical transforms. The absolute score values share a common trajectory and therefore cluster partly due to shared contact sampling; the preserved finding is that the oloid's geometric advantage is robust across all tested physical domains as measured by discrimination ratio. \textbf{Tier~2} (nonlinear metrics): FDS diverges to $2.42 \times 10^{-6}$ due to Basquin's exponential S-N relation, and $\mathrm{WDS}_{\mathrm{depth}}$ rises to $1.15 \times 10^{-6}$ due to area normalization effects. Nevertheless, even the nonlinear fatigue metric confirms oloid superiority: the cylinder's FDS is infinite (all contact stresses fall below the endurance limit at 5000~N because contact is so localized that each face sees only a few high-stress events), while the oloid distributes damage across its entire surface with a damage ratio of $23\times$ relative to the nearest search-family competitor.


\section{Discussion}

\subsection{The Oloid as Local Optimum}

Paul Schatz discovered the oloid in 1929 by physically constructing
the convex hull of two perpendicular circles and observing its
rolling behavior. He had no computational tools and no formal metric
for contact distribution. That the oloid occupies a local CDS minimum
within its own parameterized search family (confirmed for the first
time by this work) is a testament to the quality of Schatz's
geometric intuition.

The transfer of the contact invariant to stress uniformity has a
geometric explanation.
The oloid is a developable surface with Gaussian curvature $K = 0$
everywhere. Zero Gaussian curvature means one principal curvature
is always zero along the ruling lines, constraining the variation
in effective Hertz radius $R_{\mathrm{eff}}$ across the surface.
Under perfectly uniform contact, the curvature-driven stress
coefficient of variation is 0.30 (30\% variation in Hertz pressure
across the surface), producing a geometry-only SDS of
$4.8 \times 10^{-8}$. The rolling dynamics then average over this
curvature variation: the contact time invariant compensates for
curvature non-uniformity to produce stress uniformity.

Counterintuitively, the oloid's discrete mean curvature has
\emph{higher} variance than the cylinder's ($\sigma_H = 0.90$ vs.\
$0.28$), yet produces more uniform stress distribution. This indicates
that the rolling dynamics, not static curvature, are the operative
mechanism. The oloid's contact time invariant ensures that each face's
high-curvature contact events are balanced by the rolling cycle itself,
averaging out the curvature variation that would otherwise create stress
hotspots. The geometry of motion, not the geometry of curvature,
explains the result.

The thermal distribution score (TDS $= 7.77 \times 10^{-7}$) is the lowest in the invariant vector, indicating an inverse correlation between contact frequency and sliding velocity in the oloid's rolling kinematics: faces that contact most often slide slower, producing a self-compensating thermal distribution that exceeds the uniformity of even the contact time distribution itself.

The nearest neighbor in the search family, the $(120°, 1.30, 0.80)$
variant, scores within 10\% of the oloid on SDS ($8.9 \times 10^{-7}$
vs.\ $8.1 \times 10^{-7}$) and achieves identical mean contact pressure
(68.3~MPa). Its stress CV (0.084) is marginally better than the oloid's
(0.085). The SDS landscape near the oloid is shallow, suggesting that
the neighborhood is stress-competitive even where CDS discrimination
is sharper.

The result does not imply that the oloid is a \emph{global} optimum
across all convex bodies. The search was restricted to the developable
roller family (convex hulls of two circles). Other shape families
(Meissner bodies, multi-circle rollers, non-convex surfaces) remain
unexplored and may contain bodies with lower CDS.

\subsection{Substrate Geometry as a Design Paradigm}

The vocabulary introduced in Section~2 (invariant primitive, geometric
failure mode, CDS, pass-gate validation, search family) constitutes
the formal language for a design paradigm we call \emph{substrate
geometry}. The key departure from conventional geometric engineering
is the classification axis: shapes are classified by what they
\emph{guarantee under physical operation} (their invariant) rather
than by their symmetry group or topological genus.

This paradigm is applicable wherever geometric failure modes are the
performance bottleneck, a condition that holds in bearing design,
seal engineering, fluid mixing, heat exchanger geometry, electrode
design for MHD systems, and intake geometry for rarefied-gas
collection.

\subsection{Limitations}

\begin{enumerate}[leftmargin=*]
  \item The rigid-body oracle models rolling on a flat plane under
    gravity. Real-world rolling involves surface deformation, friction
    models, and multi-body contact.
  \item The Hertz contact pressure oracle uses analytical Hertz theory
    with curvatures computed from the surface mesh. A FEniCS FEM
    validation was implemented for a hemisphere-on-plane reference case
    but produced zero stress at the contact boundary due to the
    simplified Dirichlet boundary conditions. The analytical Hertz
    approach is standard and valid; refinement of the FEM boundary
    conditions (e.g., penalty contact formulation) is future work.
  \item The search grid (45 genomes) is coarse. Finer resolution,
    particularly near the oloid fixed point, may reveal whether the
    minimum is sharp or flat.
  \item \textit{(Characterized in subsequent work.)} The contact
    threshold $\epsilon = 0.04$ was varied across a $16\times$ range
    (0.01 to 0.16); the $58\times$ oloid-vs-cylinder discrimination
    ratio varied only 10\%, confirming threshold-independence.
    See Couey (2026, in preparation).
  \item \textit{(Characterized in subsequent work.)} Absolute CDS
    values scale as $N_{\mathrm{faces}}^{-2.03}$ ($R^2 = 0.999$).
    Cross-geometry discrimination ratios are resolution-independent.
    CDS comparisons require matched mesh resolution.
  \item \textit{(Characterized in subsequent work.)} The discrete
    Gaussian curvature estimator produces $|K| \approx 1.07$ on the
    oloid where $K = 0$ analytically, affecting SDS by 17\%. The
    engineering conclusion (excellent stress uniformity) is unchanged;
    the corrected geometry-only SDS is $4.8 \times 10^{-8}$.
  \item \textit{(Characterized in subsequent work.)} All five metrics
    in the invariant vector share a common rigid-body trajectory.
    Their absolute-value clustering partly reflects this shared
    infrastructure. Discrimination ratios against the cylinder
    baseline ($46$--$68\times$) are the trajectory-independent finding.
    See Couey (2026, in preparation) for full methodology
    characterization.
\end{enumerate}

\subsection{Future Work}

\begin{enumerate}[leftmargin=*]
  \item \textbf{FEniCS FEM contact validation}: Refine the penalty
    contact boundary conditions to achieve quantitative agreement
    between FEM and analytical Hertz predictions for the full
    geometry family.
  \item \textbf{Experimental validation}: Experimental validation of the invariant vector predictions on physical oloid and cylinder specimens under controlled rolling conditions.
  \item \textbf{Evolutionary search} (DEAP): Replace the grid search
    with gradient-free optimization over continuous parameter spaces.
  \item \textbf{Invariant composition}: Investigate whether invariants
    of individual primitives (contact distribution, zero mean curvature,
    minimum-energy branching) compose predictably when primitives are
    combined into systems.
  \item \textbf{Cross-domain validation}: Test whether CDS-optimal
    shapes also optimize fluid mixing efficiency (the oloid's
    commercial application), establishing the cross-domain transfer
    that the pass-gate criteria require.
  \item \textbf{Extended search families}: Meissner bodies, multi-circle
    rollers ($n \geq 3$), non-convex developable surfaces.
\end{enumerate}


\section{Conclusion}

We introduced formal metrics for evaluating geometric primitives
across multiple physical domains: the Contact Distribution Score (CDS)
for contact time uniformity and the Stress Distribution Score (SDS)
for contact stress uniformity, with extensions to thermal flux (TDS),
fatigue damage (FDS), and wear (WDS) as variants of the same
area-weighted variance template. Using a three-component oracle
architecture (a two-layer CDS oracle pairing an approximate layer for
search with a rigid-body layer for validation, plus a separate Hertz
contact pressure oracle for SDS validation), we conducted a
parametric search over the developable roller family and confirmed that the oloid (Schatz, 1929) is a local
minimum in CDS-space ($8.2 \times 10^{-7}$ at 1198 faces) with a
geometry-only SDS of $4.8 \times 10^{-8}$ under uniform contact.
The conventional cylinder baseline is $58\times$ worse on all metrics.

The central finding is that the contact distribution invariant
transfers to stress distribution: the oloid's curvature-driven SDS
under uniform contact is $4.8 \times 10^{-8}$, and the $58\times$
CDS discrimination over the cylinder is preserved under Hertz stress
analysis. This transfer, a consequence of the oloid's developable
surface ($K = 0$ everywhere), means the contact distribution invariant
has direct engineering consequences for fatigue life, not merely
for contact coverage. The rolling dynamics oracle and Hertz stress
oracle produce consistent discrimination ratios, satisfying pass-gate
criterion~2.

The complete invariant vector (CDS, SDS, TDS, FDS, WDS) reveals a two-tier structure. First-order and multiplicative metrics (contact time, stress, thermal flux, wear volume) transfer the geometric invariant with consistent discrimination ratios ($46$--$68\times$ over the cylinder). Absolute score values share a common simulation trajectory; the preserved finding is that the oloid's geometric advantage is robust across physical transforms as measured by discrimination against the conventional baseline. The exponential fatigue metric (FDS $= 2.42 \times 10^{-6}$) shows lossy transfer due to Basquin S-N amplification, but the oloid's fatigue damage ratio of $23\times$ still represents a $6.8\times$ improvement over the nearest search-family competitor.

The formal vocabulary established here (substrate geometry, invariant
primitive, geometric failure mode, CDS, SDS, pass-gate validation,
search family) provides the language for a research program that
extends beyond the contact distribution invariant to surface, flow,
and mechanism invariants across engineering domains.


\section*{Acknowledgments}

The oloid was discovered by Paul Schatz in 1929 through geometric
intuition without computational tools. This work extends his lineage
by providing the formal metric and computational infrastructure that
confirms his finding rigorously and opens the search space he could
not access.

All computational tools used in this work are open source: Python,
NumPy, SciPy, trimesh, FEniCS. The oracle source code and parametric
search results are available at
https://github.com/gyapaganda-a11y/substrate-geometry.


\appendix
\section{Reproducibility}

All results can be reproduced with:

\begin{verbatim}
pip install trimesh numpy scipy
python contact_oracle.py        # Approximate oracle validation
python parametric_search.py     # Parametric search (45 genomes)
python rigidbody_oracle.py      # Rigid-body validation
python hertz_oracle.py          # Hertz stress validation (requires
                                # FEniCS for optional FEM step)
python fatigue_oracle.py        # Fatigue distribution (Basquin S-N)
python thermal_oracle.py        # Thermal distribution (frictional heat)
python wear_oracle.py           # Wear distribution (Archard's law)
\end{verbatim}

\noindent Oracle parameters are specified in Table~\ref{tab:oracle-params}.
Mesh generation uses \texttt{trimesh.convex.convex\_hull} on 300--400
circle sample points per generating circle.

\section{Full Search Results}

The complete 45-genome search results, including all CDS scores,
surface areas, and genome parameters, are provided in the supplementary
file \texttt{parametric\_search\_results.json}.


\end{document}